# Possibilities of applying boundary functionals of random processes to nuclear safety problems

**V. V. Ryazanov**

Institute for Nuclear Research, pr. Nauki, 47 Kiev, Ukraine, e-mail: vryazan19@gmail.com

Abstract
The potential for using boundary functionals of random risk processes to solve nuclear safety problems at nuclear power plants is assessed. In certain situations (MSRs (Molten Salt Reactors), High-Temperature Gas-Cooled Reactors (HTGRs), pulverized fuel reactors, reactor startups, and accident analysis (core collapse)), neutron behavior changes significantly. Neutron clustering begins to play an important role, and the distributions characterizing neutron behavior change. The normal distribution is replaced by stable, but also limiting, distributions. Boundary functionals allow for precise calculation of the power quantile and provide a mathematical bridge between abstract directed percolation and engineering calculations of protection settings.

Article [1] examines various boundary functionals of random risk processes: first-passage time, extremums of functions, moments of return, the surplus prior to ruin, distribution of the duration of the process's stay in the lower half-plane, time the process spends above level u, and several others. They are applied to various problems: a unicyclic network with affinity A, an asymmetric random walk, non-linear diffusion, multiple diffusing particles with reversible target-binding kinetics, a two-level model, and Brownian motion and diffusion. These are general mathematical functionals. Therefore, their application to any problem is practically unlimited. In this paper, we consider the potential application of some of these boundary functionals to nuclear power plant safety issues.

In the scientific literature, the first-passage time (FPT) boundary functional is widely applied to a wide range of physical, chemical, biological, economic and other problems. In nuclear safety, the FPT functional is one of the most powerful and mathematically rigorous means of describing the achievement of a certain dangerous threshold, for example, a dangerous power level. In [2-4], it is shown that for some types of reactors (MSRs (Molten Salt Reactors), HTGRs (High-Temperature Gas-Cooled Reactors), pulverized fuel reactors) or for reactor start-up, as well as for accident analysis (core destruction), the description by the directed percolation (DP) method should be applied. In a system described by directed percolation, reaching the "dangerous level" of power $\Phi_{crit}$, FPT in the DP model is the problem of the first reaching of the threshold by a branching process. The number of descendants of each node (neutron) is a random variable k with a given probability distribution P(k). The probability P(k) obeys a stable power law

$$P(k) \sim k^{-a} \tag{1}$$

for the "effective number of descendants" - the total number of neutrons produced in all subsequent generations by one "parent" particle, under specific physical conditions. In a conventional WWER, the distribution is not a power-law. The power-law type (1) arises under conditions of strong correlation and anisotropy, characteristic of directed percolation. This is possible in three cases: a). Critical point (critical opalescence). Only at the point $k_{eff}$=1 itself (or in an infinitesimal vicinity) does the distribution of chain lengths become a power-law. b). Strong spatial inhomogeneity ("Lévy glass"). If the medium consists of empty channels and dense fuel blocks (for example, a destroyed zone or specific experimental assemblies). A neutron can fly a huge distance without collisions ("Lévy flight"). In such a medium, "descendants" are born not as a compact cloud, but as a fractal structure. c). Small number of neutrons (statistics of small samples). At startup conditions, when the core contains only a few hundred neutrons, averaging is ineffective. Research (such as in [2-4]) shows that under these conditions, the dynamics are more akin to directed percolation than diffusion.

In the classical Gaussian case, the threshold time distribution is characterized by a distribution with a narrow peak around the mean. The probability of a reactor running too fast is exponentially small. For the DP model with $a$=2, due to Lévy flights and neutron clustering, the FPT distribution acquires a heavy tail in the short-time region. There is a non-zero and significant probability that the system will reach the dangerous limit many times faster than predicted by the mean. This is called the "early ignition effect" or statistical runaway.

For a distribution function of the form (1) for the number of descendants or jump lengths with a power-law tail of the form $k^{-a}$ ($a$≈2), the characteristic function $\Phi(t)=E(e^{itX})$ for power-law tails has a non-analytic form at zero: $\ln\Phi(t) \sim -|t|^{\alpha}$. This structure $\ln\Phi(t)$ corresponds to the Lévy-Khinchin representation for stable distributions. When passing to the continuous limit (differential equation) at the macroscopic level, this term $|t|^{\alpha}$ turns into a fractional derivative with respect to space or time.

The value $a$=2 is a "watershed" for statistical moments. If $a\le 2$, then the mathematical expectation (the average number of offspring) $<k>$ diverges (tends to infinity for an infinite system size). If $2<a\le 3$, then the mean is finite, but the second moment $<k^2>$ diverges. If $a>3$, then both the mean and the variance are finite. When the variance is finite ($a>3$), the classical central limit theorem applies to the sum of such quantities, and the behavior of the system becomes "standard" in many respects. For $a>3$ (when the variance is finite), the random walk through many steps is described by the ordinary diffusion equation (the normal distribution of coordinates). For $a<3$ (including the case $a$=2), Lévy flights occur.

The application of FPT functionals to directed percolation models allows us to estimate the probability of "instantaneous" local burnouts, evaluate the sufficiency of the response speed of protection systems, and take into account the stochastic nature of the launch, which is ignored by deterministic codes (such as KORSAR or RELAP).

Article [1] shifts the focus from simple average values to the full statistics of boundary functionals of stochastic processes, which includes FPT [1]. This article examines FPT. However, [1] also derives a number of other boundaries functionals that are no less significant for NPP safety.

For processes described by stable Levy laws (index $\alpha=a-1\approx 1$), the probability density of the time of first reaching the threshold L has the asymptotics:

$$P(t_{FPT}) \sim t^{-(1+\alpha/z))}, \qquad (2)$$

where z is the dynamic critical index of directed percolation. The average time to threshold $<t_{FPT}>$ may formally diverge or have a huge variance. Therefore, the "mean time to failure" will have no significance for safety, since the actual spread of times to reach the critical level covers several orders of magnitude.

How this can be applied to WWER safety assessments. When analyzing reactor startup or operation at the MCL (minimum control level), the FPT functionality is used to solve three problems: a). Adjustment of the EP (Emergency Protection) setpoints. The period (rate of increase) setpoint should be configured to "cut off" fast spikes from the tail of the FPT distribution, but not to cause false alarms due to normal statistical noise. b). Equipment dead time. If the FPT distribution has a heavy tail in the region of small times, there is a risk that the system acceleration time will be less than the protection response time (rod drop time + information management system (IMS) logic time). c). Probabilistic safety analysis (PSA). Instead of a single accident scenario, an FPT distribution is constructed. Safety is considered ensured if the probability integral $P(t_{FPT}<t_{action})$ is negligible (where $t_{action}$ is the system response time).

The effect of "truncation" (finiteness of the system). The finite size of the reactor (L) "truncates" the Lévy flights. For FPT, this means that at very long times, the distribution will still become exponential. However, dangerous limits are usually reached at short times, where the truncation has not yet kicked in. That is, in the risk zone (rapid accelerations), the reactor behaves precisely like a "pure" directed percolation system, with all its anomalies.

Let us consider the dynamics of the local neutron flux Φ(t) near the critical point. In the representation of directed percolation with a stability index α (where $\alpha=a-1$), the evolution equation has the form of a fractional stochastic equation:

$$\partial^{\gamma}\Phi / \partial t^{\gamma} = D_{\alpha}(-\Delta^{2})^{\alpha/2}\Phi + \eta(r,t)\,, \tag{3}$$

where ($0<\gamma<1$), $\eta(r,t)$ is non-Gaussian noise with heavy tails, characteristic of branching processes. In this case, the dynamic index becomes equal to: $z=z_{dyn}=2\alpha/\gamma$.

Let $\Phi_{crit}$ be the critical power limit (e.g., the local boiling threshold or the emergency shutdown setpoint). The time to first reach $T_{FPT}$ is defined as:

$$T_{FPT}=\inf\{t>0:\ \Phi(t)\geq\Phi_{cri}\}. \tag{4}$$

In contrast to classical diffusion, where the asymptotic behavior of $P(T_{FPT})$ is exponential, in the directed percolation regime with Lévy statistics the distribution has an algebraic tail:

$$P(T_{FPT}>t)\sim t^{-\theta}\,, \tag{5}$$

where θ is the critical "survival" index in DP theory. For $a=2$, this index determines the high probability of abnormally rapid process realizations.

The impact of truncation on safety. In a real WWER reactor, the finite size L introduces a truncation parameter $\lambda\sim 1/L$. This modifies the probability density f(t) of the time to reach:

$$f(t)\approx t^{-(1+\theta)}\exp(-\Gamma t). \tag{6}$$

At short times ($t<<1/\Gamma$): the percolation nature dominates. The probability of fast transient is significantly higher than in Gaussian calculations. At long times ($t>>1/\Gamma$): the reactor geometry "dampens" correlations, returning the system to exponential risk decay. To ensure safety, the protection response time $t_{prot}$ must satisfy the condition:

$$\int_{0}^{t_{prot}} f(t)dt < P_{acc}\,, \tag{7}$$

where $P_{acc}$ is the maximum permissible accident probability (e.g., $10^{-7}$ per year). When using the directed percolation model ($a=2$), the probability integral in the short-time domain (0, $t_{prot}$) is orders of magnitude larger than in the classical model.

Consequently, traditional calculations of the EP response time may contain a hidden safety flaw, as they fail to account for the possibility of the "instantaneous" formation of a percolation cluster, bypassing average diffusion rates. The nature of the process is such that the startup and operation of a VVER reactor at low power levels involves dynamic, directed percolation, not steady-state diffusion.

In [1], the roots of the Lundberg equation (8) appear in the expressions for the boundary functionals;

$$G(k)=s, \tag{8}$$

where G(k) is the logarithm of the characteristic function (more precisely, the generating function of moments $M(k)=E[e^{kX}]$, obtained from the characteristic function by analytical continuation and replacement $iq \to k$), and s is the Laplace transform parameter for time. For the case $a=2$ (critical directed percolation), the Lévy-Khinchine integral is calculated analytically, resulting in a logarithmic structure for the logarithm of the characteristic function $\Psi(q) = -\sigma|q|(1 + i\beta(2/\pi)\,\mathrm{sgn}(q)\ln|q|) + i\mu q$. This explicit form, corresponding to the critical $\alpha=1$, transforms the Lundberg equation into a logarithmically "tightened" asymptotic, explaining the anomalously high-power peaks. The Lundberg equation for the case of stable Lévy distributions ($a \le 2$, $\alpha \le 1$) is transcendental. Its solution k(s) determines the poles in the Laplace images for boundary functionals (FPT, maximum of a random process, etc.). In the case of directed percolation with heavy tails, we do not have a simple quadratic solution, as in ordinary diffusion ($Dk^2=s \to k=\sqrt{s/D}$).

Adding the Doppler effect (negative feedback) to the Lundberg equation for directed percolation is a crucial step. Physically, this means that the system gains a "restoring force": the higher the power Φ, the more strongly the Doppler effect pulls it downward. Mathematically, the logarithm of the characteristic function G(k) (in terms of [1]) introduces a linear term -vk, where v is the effective reactivity suppression rate.

Formation of the Lundberg equation with the Doppler effect. For a stable Levy process (index $\alpha \approx 1$, which corresponds to $a=2$) taking into account the "truncation" by geometry ($L \sim 1/\lambda$) and the Doppler effect (v), the Lundberg equation takes the form: $vk - \sigma((\lambda-k)^{\alpha} - \lambda^{\alpha}) = s$, where: s is the Laplace parameter (conjugate of time T). v is the Doppler coefficient (velocity of reversion to the mean). σ is the intensity of neutron flares (percolation parameter), λ is the geometric truncation (reactor boundaries). In the case of most interest from a safety standpoint ($\alpha \approx 1$), the equation becomes practically linear: $vk + \sigma k = s \to k(s) = s/(v+\sigma)$. Stabilization occurs: the Doppler effect (v) sums with the percolation attenuation (σ). This increases the denominator, decreasing the root k(s).

A linear risk is observed: even with Doppler, the dependence remains linear in s (rather than root-valued $\sqrt{s}$, as in diffusion). This confirms that stochastic spikes in a heterogeneous medium occur faster than predicted by conventional kinetics.

Let's consider other boundary functionals. The physical meaning of the maximum in a nuclear reactor. If the FPT answers the question, "When will we reach the limit?", then the maximum functional on the interval T answers the question, "What was the peak local power?" Thermal inertia may be important. Fuel rods can withstand a short-term power surge if its integral or peak does not exceed the physical failure threshold.

What is the connection with DP ($a=2$)? In processes with Lévy statistics, the process maximum grows with time much faster than in Gaussian systems. For $a=2$, the mathematical expectation of the maximum can behave as $\langle M_T \rangle \sim T^{1/\alpha}$, which, at $\alpha \approx 1$, yields a linear increase in peak emissions with observation time.

Application of other functionals from article [1]. In addition to the maximum, the paper also considers other boundary functionals that are ideally suited to reactor physics.

Occupation Time. Definition: the total time the flux Φ(t) was above the safe threshold. Application: assessing fuel rod cladding degradation. A single short-term "spike" (maximum) may not melt the steel, but a series of percolation bursts ("spotting") can lead to accumulation of fatigue or corrosion.

Level Crossings. Meaning: how often a process “break” the setpoint. Application: adjusting the control system sensitivity. If the level crossing rate is abnormally high in DP mode ($a$=2), this creates a problem of "false alarms" and wear on the rod drives.

Overshoot. Essentially, it's the magnitude of the flux's overshoot at the moment of its first impact. Application: in Lévy statistics (Lévy flights), the overshoot is always positive and can be large. Unlike continuous diffusion, where the process "touches" the threshold, in directed percolation its "breaks" through it with a jump. This is critical for estimating the margin before melting.

Formalization of Extreme Value Statistics (EVS) for WWER reactors. Methods [1] allow us to relate the canonical Levy-Khinchin form to extreme value distributions (Gumbel or Fréchet type). For $a$=2, the distribution of power surge maxima in the reactor will belong to the Fréchet class (heavy tails). This means that "record" power surges are statistically inevitable. Safety: protection design should be based not on the "average maximum," but on the quantiles of the extreme value distribution obtained through boundary functionals. Thus, work [1] provides a tool for quantitatively assessing the magnitude of surges. If FPT is the "time to alarm", then the maximum and residence time functionals represent the "cost of damage" from the stochastic nature of directed percolation.

Statistics of extreme emissions in the directed percolation regime. Maximum functional. We determine the maximum local neutron flux $M_T$ over the time interval [0, T]:

$$M_T=\max_{0\leq t\leq T}\Phi(t). \qquad (9)$$

In classical diffusion (Gaussian), the tail of the maximum distribution decays exponentially. However, for processes with Lévy statistics ($a$=2, stability index $\alpha=a-1=1$), the distribution of the maximum $P(M_T>x)$ obeys a power law.

Fréchet Distribution. According to limit theorems for stable processes, the distribution of the maximum sum (or chain) of divisions in the directed percolation regime at $T\to\infty$ converges to the Fréchet distribution:

$$P(M_T\leq x)\approx\exp(CTx^{-\alpha}), \qquad (10)$$

where $\alpha\approx1$ is an indicator determined by the index of directed percolation, C is a scaling parameter associated with the intensity of fission and the canonical Levy-Khinchin form, T is the observation time (the duration of the start-up or operation at the MCL).

Mathematical assessment of the probability of a catastrophic release. From the Fréchet formula, it follows that the probability of the peak power exceeding the dangerous threshold $\Phi_{crit}$ decreases extremely slowly:

$$P(M_T>\Phi_{crit})\approx CT\Phi^{-\alpha}{}_{crit}. \qquad (11)$$

Physical consequence for $a$=2 ($\alpha$=1). Linear risk growth: the probability of encountering an abnormally high release increases linearly with the reactor's operating time at low power. Absence of a "safe average": the mathematical expectation of the maximum $\langle M_T\rangle$ for the pure Lévy distribution at $\alpha$=1 formally diverges (in reality, it is limited by the reactor geometry L). This means that the "average release" is not a reliable safety measure.

Application in Safety Analysis. The use of boundary functionals [1] allows us to reconsider the concept of "margin before melting".

Overshoot: since the DP model divides in "jumps" (clusters), at the moment the protection is triggered, the power does not simply touch the threshold, but "breaks through" it by a value of $\Delta\Phi=M_T-\Phi_{crit}$. For $a$=2, this value can be comparable to the threshold itself.

Designing for "records": reactor protection should be calculated not for average fluctuations, but for extreme values (Extreme Value Theory). Boundary functionals allow for the precise calculation of a power quantile that will not be exceeded with a probability of, for example, 0.9999.

Summary. The work [1] provides a mathematical bridge between abstract directed percolation and engineering calculations of protection settings. While the FPT determines how quickly the system will respond, the maximum functional determines the magnitude of the shock (thermal or neutron) a fuel element must withstand during this stochastic flare.

Key conclusion: at $a$=2, "rare" events become statistically significant. VVER safety at low power requires taking into account the statistics of records, not just average values and variances.

**References**

1. V. V. Ryazanov (2025) Application of boundary functionals of random processes in statistical physics. Physical Review E, 111(2), 024115. doi:10.1103/PhysRevE.111.024115.

2. B. Dechenaux, T. Delcambre, E. Dumonteil (2022) Percolation properties of the neutron population in nuclear reactors. Physical Review E, 106, 064126.

3. E. Dumonteil et al. (2014) Particle clustering in Monte Carlo criticality simulations. Annals of Nuclear Energy, Vol. 63 Pages 612-618.

4. E. Dumonteil, R. Bahran, T. Cutler, B. Dechenaux, T. Grove, J. Hutchinson, G. McKenzie, A. McSpaden, W. Monange, M. Nelson, N. Thompson, and A. Zoia (2021) Patchy nuclear chain reactions. Communications Physics, 2021, 4 (1), pp.151. https://doi.org/10.1038/s42005-021-00654-9.